\documentclass[10pt, twocolumn, conference]{IEEEtran} 
\usepackage[utf8]{inputenc}
\usepackage{caption}
\usepackage{multicol}
\usepackage{multirow}
\usepackage{amsmath,amssymb,amsfonts}
\usepackage{algorithmic}
\usepackage{graphicx}
\usepackage{textcomp}
\usepackage[dvipsnames]{xcolor}
\usepackage[utf8]{inputenc}
\usepackage{caption}
\usepackage{times,amsfonts,graphicx}
\usepackage[perpage,symbol]{footmisc}
\usepackage[english]{babel}
\usepackage{amsmath, bm}
\usepackage{pbox, booktabs, rotating}
\usepackage{array, makecell, lipsum}
\usepackage{enumerate}
\usepackage{threeparttable}
\usepackage{multirow,hyphenat,balance}
\usepackage{placeins}
\usepackage[compress]{cite}
\usepackage{float,color}
\usepackage{ragged2e}
\usepackage{enumerate}
\usepackage{etoolbox}
\usepackage{paralist}
\usepackage{cases}
\usepackage{url}
\usepackage{framed}
\usepackage{flushend}
\usepackage{amsmath}
\usepackage{amsthm}
\usepackage{amsthm}
\usepackage[english]{babel}
\usepackage{amssymb}
\usepackage{subfig}
\usepackage[linesnumbered,ruled]{algorithm2e}
\let\oldnl\nl
\newcommand{\nonl}{\renewcommand{\nl}{\let\nl\oldnl}}
\usepackage{pifont}
\usepackage{graphicx}
\usepackage{subcaption}
\usepackage{commath}
\usepackage{flushend} 
\usepackage{soul}
\usepackage{hyperref}
\usepackage{cleveref}

\usepackage{mathtools}

\newcolumntype{P}[1]{>{\centering\arraybackslash}p{#1}}
\newcolumntype{M}[1]{>{\centering\arraybackslash}m{#1}}

\DeclareCaptionLabelFormat{nospace}{#1#2}
\title{Reward-based Blockchain Infrastructure for 3D IC Supply Chain Provenance}

\author{
    \IEEEauthorblockN{
        Sulyab Thottungal Valapu$^\dagger$,
        Aritri Saha$^*$,
        Bhaskar Krishnamachari$^\ddagger$, 
        Vivek Menon$^{**}$, and
        Ujjwal Guin$^*$} 
\IEEEauthorblockA{
    $^\dagger$Dept. of Computer Science, University of Southern California\\
    $^*$Dept. of Electrical and Computer Engineering, Auburn University\\
    $^\ddagger$Dept. of Electrical and Computer Engineering, University of Southern California\\
    $^{**}$National Reconnaissance Office \\
    \text{Emails:
        \{thottung, bkrishna\}@usc.edu},
        \{aritri.saha, ujjwal.guin\}@auburn.edu, and
        menonviv@nro.mil
    }
}

\begin{document}

\maketitle

\thispagestyle{plain}
\pagestyle{plain}

\begin{abstract}
In response to the growing demand for enhanced performance and power efficiency, the semiconductor industry has witnessed a paradigm shift toward heterogeneous integration, giving rise to 2.5D/3D chips.
These chips incorporate diverse chiplets, manufactured globally and integrated into a single chip.
Securing these complex 2.5D/3D integrated circuits (ICs) presents a formidable challenge due to inherent trust issues within the semiconductor supply chain.
Chiplets produced in untrusted locations may be susceptible to tampering, introducing malicious circuits that could compromise sensitive information.
This paper introduces an innovative approach that leverages blockchain technology to establish traceability for ICs and chiplets throughout the supply chain.
Given that chiplet manufacturers are dispersed globally and may operate within different blockchain consortiums, ensuring the integrity of data within each blockchain ledger becomes imperative.
To address this, we propose a novel dual-layer approach for establishing distributed trust across diverse blockchain ledgers.
The lower layer comprises of a blockchain-based framework for IC supply chain provenance that enables transactions between blockchain instances run by different consortiums, making it possible to trace the complete provenance DAG of each IC.
The upper layer implements a multi-chain reputation scheme that assigns reputation scores to entities while specifically accounting for high-risk transactions that cross-blockchain trust zones.
This approach enhances the credibility of the blockchain data, mitigating potential risks associated with the use of multiple consortiums and ensuring a robust foundation for securing 2.5D/3D ICs in the evolving landscape of heterogeneous integration.
\end{abstract}

\vspace{5px}

\section{Introduction}
The increasing demands for computational power in high-performance computing (HPC), data centers, cloud computing, and machine learning have surpassed the capabilities of the current System-on-Chip (SoC) paradigm. Meeting the challenges posed by issues such as latency, power consumption, throughput, and fabrication yield requires innovative solutions. In response to these demands, heterogeneous integration (HI) with 2.5D/3D packaging has emerged as a transformative strategy. These groundbreaking approaches involve both horizontal and vertical stacking of multiple dies within a single package or chip. The departure from the monolithic IC architecture seen in traditional SoC designs represents a paradigm shift in the semiconductor industry, with HI and 2.5D/3D packaging offering innovative solutions to the challenges of latency, power consumption, throughput, and IC fabrication yield~\cite{vashistha2022toshi, tsmc2020blog3dfabric}. The active exploration of these technologies is evident in the ongoing research and development efforts of various industry players. Notable initiatives include TSMC's 3DFabric$^{\text{TM}}$, which focuses on 3D silicon stacking and advanced packaging technologies~\cite{tsmc3dfabric-page}, and Samsung's 3D-TSV (12 layers) DRAM Chip~\cite{samsung-2019-3dmemory}. The establishment of ubiquitous interconnect standards, e.g., Universal Chiplet Interconnect Express (UCIe)~\cite{UCIe}, at the package level has facilitated die-to-die communication in these advanced packaging approaches.

Unfortunately, the globalization of the semiconductor supply chain exposes critical vulnerabilities, including untrusted electronic products, counterfeit ICs, and devices compromised with hardware Trojans. These threats stem from malicious third-party IP vendors, untrusted manufacturing facilities, rogue distributors, and various untrusted entities in the supply chain. Bloomberg notably reported hardware hacks in 2018 and 2021, involving covertly placed extra tiny chips on boards capable of compromising sensitive data from US companies~\cite{bloomberg-big-hack-2018, bloomberg-long-hack-2021}. While the reported hacks targeted pre-heterogeneous integration (HI) hardware, the potential for adversaries to execute similar attacks by incorporating malicious die(s) or chiplets inside 2.5D/3D packages persist. Compromised hardware, whether unsecured or unreliable, creates opportunities for mounting software attacks. This includes exploiting firmware and software vulnerabilities to gain unauthorized access to the device or system by bypassing existing security measures implemented at the software level.

The security community continually devises diverse solutions to combat these intricate hardware security threats~\cite{bhunia2018hardware, tehranipoor2015counterfeit}. Unfortunately, the universal adoption of these solutions remains elusive. Firstly, establishing integrity is paramount to ensure that all participating entities can rely on the supply chain. Secondly, providing incentives becomes imperative to encourage entities in the supply chain to implement robust security measures. Lastly, the assessment of trust and security measures must be based on observable and quantifiable metrics. For the industry to effectively address these challenges, there is a pressing need for the adoption of observability and traceability in the supply chain, aligning these measures with business interests.

The widespread interest in incorporating blockchain to ensure supply chain provenance is evident across academia, industry, and government. This attention is driven by the inherent properties and features of blockchain, which have the potential to improve the traceability, transparency, and reliability of the supply chain~\cite{zhong2023blockchain, cui2019blockchainProvenance, xu2019electronics, crosby2016blockchain, pilkington201611, guin2018ensuring, islam2019enabling}. However, while numerous challenges within the realm of supply chain have been addressed, the adoption of these solutions has yet to be accepted for electronic parts. A crucial question persists regarding the motivation of an entity in the electronics supply chain to participate in these blockchain initiatives. The underlying incentive or reward-based mechanism for motivating participation and ensuring trustworthiness in the blockchain framework still remains widely unexplored.
The key challenge lies in establishing trust between blockchains run by separate, autonomous \emph{consortiums}.
Areas of concern include data quality (transactions may be falsified or omitted), lack of transparency into the KYC process (sybil identities may exist), and trustworthiness of miners.
Addressing these concerns is essential for relevant entities to perceive the added overhead of blockchain-based supply chain provenance as beneficial and worthwhile.

This paper provides a comprehensive solution to these challenges by presenting a reward-based blockchain system designed to facilitate the seamless movement of chiplets and ICs across interconnected consortium blockchains.
The proposed infrastructure introduces a novel approach, leveraging a reputation-based system to instill trustworthiness in the chain and serve as a motivation for organizations to take part in our framework.
Notably, when the nodes consistently exhibit trustworthy behavior, they are rewarded which reinforces our previous point of incentive-based blockchain architecture.
Furthermore, our architecture extends beyond the individual consortium chains, establishing a solid foundation of trust that spans across interconnected networks, whether the chiplet fabrication is located on-shore(trusted) or off-shore(untrusted).

\subsection*{\bf Contributions.}
\begin{itemize}
    \item \textit{Reputation Scheme}:
    We propose a novel multi-chain reputation scheme specifically tailored for real-world IC supply chains. Our scheme is designed to operate with independent blockchain consortiums, reflecting real-world business and administrative trust domains where a unified trust metric may not be acceptable due to a lack of mutual trust. In our dual-layer approach, the lower layer enables IC supply chain provenance by facilitating transactions between consortium-run blockchains, while the upper layer dynamically assigns reputation scores to entities using an additive increase and multiplicative decrease approach. \textit{\textbf{To the best of our knowledge, we are the first to introduce a comprehensive method for assigning reputation scores to blockchain members by rigorously evaluating their performance as they engage in transactions across various stages of the semiconductor supply chain.}} This approach not only enhances the accountability of individual participants but also strengthens the overall resilience and credibility of blockchain-based systems in the semiconductor supply chain by promoting ethical behavior and reducing the risk of malicious activities. Our scheme, therefore, represents a significant advancement in managing trust and reputation within decentralized environments.

    \item \textit{Broader Blockchain Adoption:} One of the significant challenges hindering the adoption of blockchain in the semiconductor supply chain is the lack of sufficient incentives for participating entities. Our proposed blockchain infrastructure is specifically designed to address this issue by offering tangible rewards to members who maintain high reputation scores. Entities with a strong reputation will benefit from easier and more efficient sales processes because their trustworthiness is recognized across the network, and thus reducing transaction friction. The system is not only scalable, as it allows seamless growth as more members join, but it is also structured to be user-friendly, making integration into the consortium straightforward for new participants. This combination of incentives, ease of entry, and scalability ensures that the blockchain network remains robust, efficient, and attractive to a broad spectrum of stakeholders.

    \item \textit{Simulation Platform for Semiconductor Supply Chain:} We have developed a comprehensive simulation platform specifically designed for the complex semiconductor supply chain. \textit{\textbf{To the best of our knowledge, this is the first instance where a directed acyclic graph (DAG) has been proposed as a model for representing the intricate dependencies and processes within the supply chain.}} By introducing this novel approach, we aim to provide the research community with a powerful tool to facilitate a deeper understanding of semiconductor supply chain dynamics. We believe this platform will enable more accurate modeling, offer opportunities for optimizing various supply chain-related applications, and allow researchers to test ideas without needing to implement them in real-world supply chains, which is often impractical in academic settings.
\end{itemize}    

The rest of the paper is organized as follows. \Cref{sec:related-work} covers prior works on using blockchain for supply chain provenance. In~\Cref{sec:blockchain}, we present our proposed blockchain architecture for supply chain provenance, including the chiplet ecosystem. We present our proposed reputation scheme in~\Cref{sec:reputation}. We analyze the effectiveness of our proposed framework in~\Cref{sec:eval}. In~\Cref{sec:discussions}, we discuss the strengths and limitations of our proposed scheme and highlight potential areas for future research. Finally, we conclude the paper in~\Cref{sec:conclusion}.

\section{Related Work} \label{sec:related-work}

Blockchain-based reputation schemes have been widely explored in research for diverse applications ranging from e-commerce~\cite{sun2023rtchain} and crowdsourced ratings/reviews~\cite{thottungal_valapu_darsan_2023} to quality control for IoT data~\cite{dedeoglu2020trust} and reputation-aware routing for satellite constellations~\cite{clark2020blockchain}.
Reputation calculation methods are typically tailored to suit the properties and requirements of specific applications, such as scale of deployment, computational capacity of nodes, and the availability and trustworthiness of reputation ``signals''.
Bellini et al.~\cite{bellini2020blockchain} provides an excellent survey on the subject.

Blockchain research has been actively carried out across different supply chains, such as pharmaceutical industries\cite{abbas2020blockchain,abdallah2023blockchain}, agriculture\cite{bhat2021agriculture} and the clothing industry~\cite{guo2020applications, wang2020blockchain}, addressing both security and management challenges to streamline the process of the movement of assets across the supply chain. Few research on blockchains has stepped foot onto the realm of vendor-managed inventory~\cite{omar2020enhancing, dasaklis2019improving}, where the vendor observes the end users' supply and sales and makes decisions regarding stocking of supplies accordingly.

Over the years, much work has been extended to implement a blockchain framework to enhance security in IoT devices and the semiconductor supply chain. Guin et al.~\cite{guin2018ensuring} proposed a blockchain-enabled framework that assured the authenticity of devices using an unclonable identifier (ID) generated from an SRAM PUF. Cui et al. proposed a confirmation-based ownership transfer of devices across the chain using Hyperledger fabric~\cite{cui2019blockchainProvenance}, while Zhong et al. proposed a modular framework, utilizing many of the functions proposed in~\cite{cui2019blockchainProvenance}, for protection of trade secrets across different supply chain lifecycles. Additionally, to keep up with Moore's Law, the introduction of 3D ICs has exposed the supply chain to several attacks at different stages of die manufacturing and assembly due to the nature of decentralized production. Calzada et al. have proposed a blockchain framework to establish authentication of a SiP throughout its lifecycle using a Chiplet hardware Security Module~\cite{calzada2023heterogeneous}. 

While our paper draws inspiration from the three research areas summarized above, to the best of our knowledge, we are the first to introduce a blockchain-based reputation framework for semiconductor supply chains to support inter-operation between blockchains controlled by independent consortiums.

\section{Blockchain for Supply Chain Provenance} \label{sec:blockchain}
The widespread interest in incorporating blockchain to ensure supply chain provenance is evident across academia, industry, and government. This attention is driven by the inherent properties and features of blockchain, which have the potential to greatly improve the traceability, transparency, and reliability of the supply chain~\cite{zhong2023blockchain, cui2019blockchainProvenance, xu2019electronics, crosby2016blockchain, pilkington201611, guin2018ensuring, islam2019enabling}. Guin et al. presented a traceability framework that utilizes a layered, scalable, and permissioned blockchain~\cite{zhong2023blockchain, cui2019blockchainProvenance}. Tailored for versatility across diverse industries, this framework automates the tracking and traceability of electronic parts and systems from the design phase to end-of-life. The proposed infrastructure guarantees not only traceability but also safeguards privacy, protects trade secrets, and ensures the integrity of the electronic parts. In this section, we delve into a comprehensive analysis outlining the steps and considerations involved in constructing a provenance framework for chiplets manufactured on a global scale.

\begin{table}[ht]
    \centering
    \caption{Participating entities in the blockchain-based traceability framework.}
    \begin{tabular}{P{0.6in}|p{2.5in}}
    \hline
    
       \textbf{Entity}  &  \textbf{Description}\\ \hline \hline
       Chiplet \text{Manufacturer}  & Fabricates chiplets. Chiplet manufacturers can directly send chiplets to the IC manufacturers. However, they can send chiplets to their authorized distributors as well.  \\\hline
       
       Chiplet Distributor & Distributes chiplets and can be untrusted.   \\ \hline       
       IC \text{Manufacturer}/ Foundry/Fab  & Owns a foundry and fabricates ICs. Most IC design houses typically do not possess their own foundry but instead contract the fabrication of ICs to a fabrication facility (fab). Note that design houses can be excluded as they are not directly involved in handling chiplets. \\\hline 
       
       IC Distributor  & Distributes ICs, and can be untrusted.\\\hline
       
       System Integrator (\textit{SI})  & Designs electronic systems. \textit{SIs} use ICs to build complex systems. \\\hline  
       
       End User (\textit{EU}) & Plays a crucial role in utilizing ICs for the construction and maintenance of complex systems. \textit{SIs} can function as a category of \textit{EUs}. Note that \textit{EUs} may not actively participate in the blockchain framework and can serve as an off-chain entity. \\\hline
    \end{tabular} 
    \label{tab:entities}
\end{table}

\Cref{tab:entities} summarizes the various entities and their roles in the electronics supply chain. The above-mentioned entities represent the design, manufacturing, integration, assembly, and distribution phases in the supply chain. We typically consider end users and system integrators as trusted entities, while distributors may fall into both trusted and untrusted categories. Adversaries have the potential to create cloned devices, integrate recycled or used devices, or introduce tampered devices with hardware Trojans or malware into ICs.

\begin{figure*}[tb]
    \centering
    \includegraphics[width=1\textwidth]{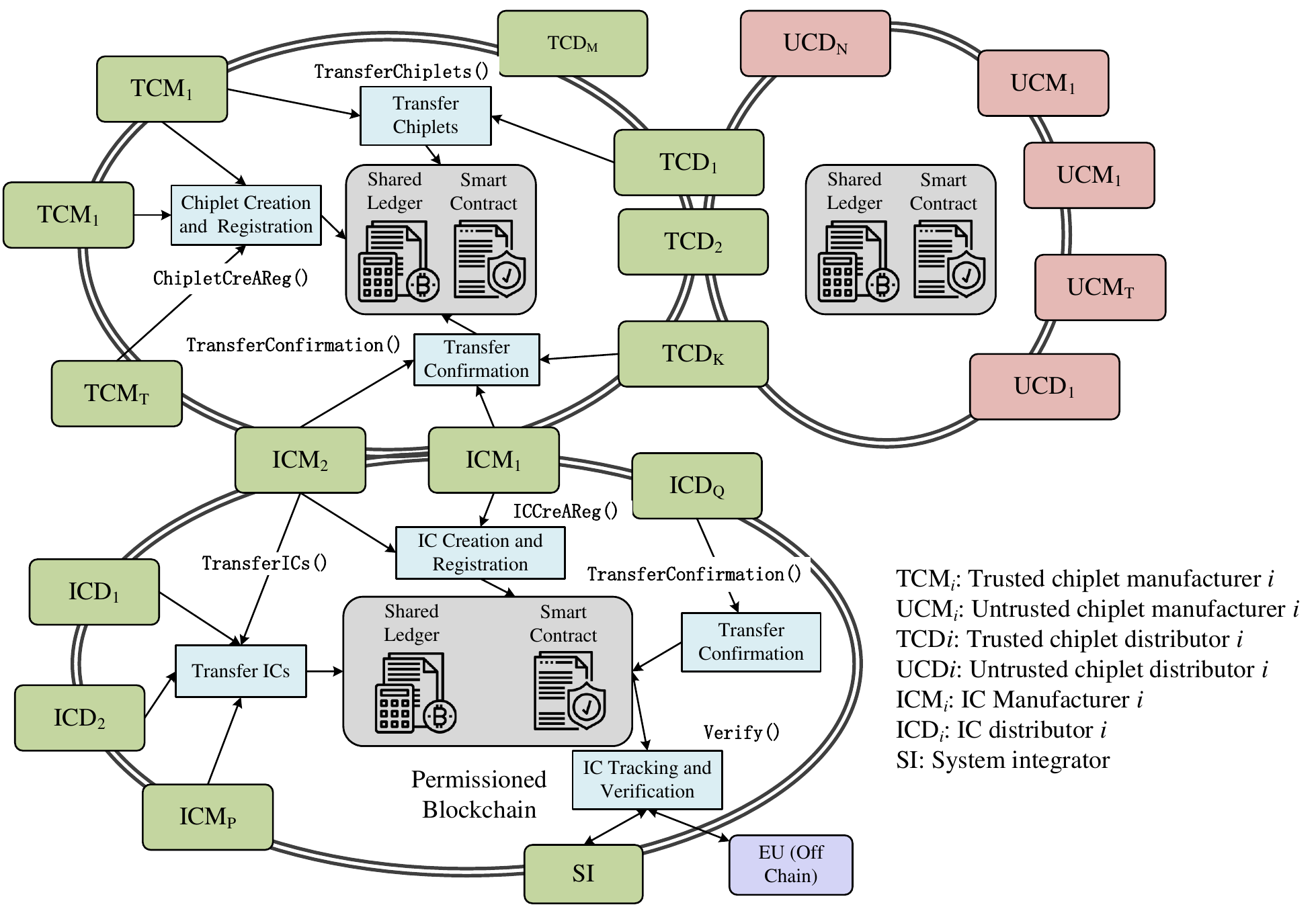}
    \caption{Proposed blockchain-based traceability framework for chiplet and IC supply chain provenance. \vspace{-10px}
    }
    \label{fig:blockchain}
\end{figure*}

\Cref{fig:blockchain} shows the overall provenance of chiplet and IC ecosystem. Chiplet and IC manufacturers have the exclusive authority to register device types (chiplets or ICs) on the blockchain. Each device type must undergo separate registration within the blockchain, a process facilitated by the use of smart contracts. IC manufacturers utilize the $\mathtt{ICCreAReg()}$ function to formally register new IC types. For instance, TSMC may create an entry for the Qualcomm Snapdragon 800 Processor type to track all the processors manufactured. On the other hand, chiplet manufacturers are restricted to registering only their chiplet types using $\mathtt{ChipletCreAReg()}$ function. The creation and registration of these device types are permanently recorded on the blockchain and visible to all downstream chain members and major identified upstream participants who require notifications. It is important to note that only chiplet and IC manufacturers possess the capability to register device types within the blockchain system. Other participants, such as distributors, system integrators, or end users, are not permitted to register device types due to the specified blockchain policy.

After registering device types in the blockchain, chiplet or IC manufacturers proceed to register individual devices for traceability. To ensure traceability, a unique device ID is essential, achievable through the integration of an ECID~\cite{CiscoECIDBillEklow}, PUF~\cite{gassend2002silicon, suh2007physical, guajardo2007fpga, wang2018ramprate}, or another unique identification method. Rather than directly placing the ID in the blockchain, we advocate storing the hash of the ID for enhanced security. This approach prevents the determination of the original ID unless one possesses the actual devices. While all members are aware of the uploaded hashes and the number of entries, the use of hashing ensures that actual IDs are not disclosed.

It is imperative to record the transfer of devices (e.g., chiplets and ICs) among different entities in the supply chain to ensure traceability. This can be achieved through the utilization of $\mathtt{TransferICs()}$ and $\mathtt{TransferChiplets()}$. Consider the following example: Entity X intends to transfer a quantity $N_1$ of chiplets to Entity Y from its possession of $N$ chiplets. Entity X initiates a device transfer transaction function with $\mathtt{TransferChiplets(chiplet\_name, N_1, \{ID_{N1}\},\{SP_{N1}\} Y)}$. It is important to note that, depending on the implementation, transferring $N_1$ entries in the blockchain may involve $N_1$ transactions, each denoting the transfer of one chiplet or IC, containing one hashed ID, or alternatively, a single transaction (or several) could include all the hashed IDs. 
It is permissible for manufacturers and distributors to initiate transactions for device transfers, provided they possess a specific quantity of devices. However, it is crucial to understand that the actual ownership of the device declared to be transferred in the device transfer transaction does not transition to the new owner until a confirmation transaction is received to address in-transit thefts.

When an entity (such as a manufacturer or distributor) delivers a specified quantity of electronic devices to a new owner, the recipient must initiate a confirmation transaction. The device transfer process remains incomplete and unverified until the confirmation of the transfer is officially acknowledged. The trace and ownership of the device are transferred within the smart contract only after the confirmation has been successfully processed. The confirmation transaction function, denoted as $\mathtt{TransferConfirmation()}$, receives the following inputs: $\mathtt{(chiplet\_name/IC\_name, N_1, \{ID_{N1}\})}$. The invoking of the confirmation process, indicates a successful transaction.

Upon physically receiving a device (IC/chiplet), participants are obligated to verify its identity (ID) using $\mathtt{Verify()}$, which is present (hashed) in the blockchain. The verification process necessitates retrieving the unique device ID, accessible through the JTAG interface~\cite{JTAG} or similar methods. If the ID is not found in the system, a flag will be raised, marking the device as suspicious. It is important to note that the verification and tracking procedure does not modify the data stored in the blockchain. Consequently, no actual transaction occurs, rendering the entire process highly efficient.

\section{Multi-Chain Reputation Scheme} \label{sec:reputation}
This section presents a comprehensive design of a multi-chain reputation scheme, which builds upon the blockchain traceability framework introduced earlier in~\Cref{sec:blockchain}. To begin with, it is crucial to recognize that trust and reputation are inherently subjective concepts, which differ significantly from the objective nature of provenance. For instance, one blockchain ($B_X$), might perceive another blockchain ($B_Y$) as untrustworthy based on its criteria and experiences, while blockchain $B_Y$ could hold a similarly skeptical view of blockchain $B_X$. This subjectivity underscores the complexity of establishing a universally accepted reputation system across multiple blockchains. Given this, our subsequent discussion on reputation is explicitly framed from the perspective of the blockchain of which the System Integrator (\textit{SI}) or End User (\textit{EU}) is a part. This approach allows us to develop a reputation scheme that is adaptable to the diverse and decentralized nature of multi-chain environments, ensuring that trust can be managed in a way that reflects the unique characteristics and requirements of each blockchain network involved.

\begin{figure}[tb]
    \centering
    \includegraphics[width=\linewidth]{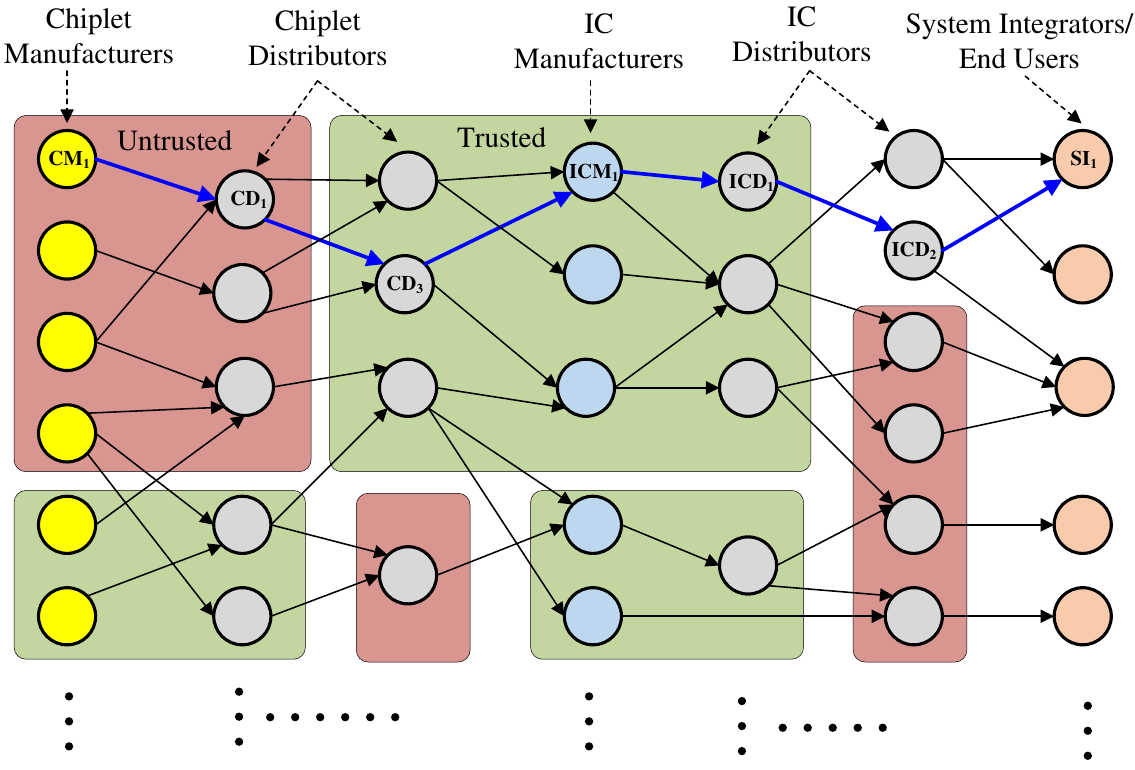}
    \caption{Semiconductor supply chain represented as a DAG.} \vspace{-10px}
    \label{fig:DAG}
\end{figure}

\subsection{Directed Acyclic Graph (DAG) of Supply Chain} \label{subsec:DAG}
We have proposed modeling the semiconductor supply chain with a directed acyclic graph (DAG) to represent the intricate dependencies within the supply chain for computing the trust score, as shown in~\Cref{fig:DAG}. The arrows represent the connection between the supply chain entities as parts travel from seller to buyer. For example, a chiplet travels from its manufacturer ($CM_1$) through many distributors ($CD_1$, and $CD_3$), before finally reaching an IC manufacturer ($ICM_1$), where it is consumed into an IC. The chips then travel across different distributors ($ICD_1$ and $ICD_2$) and finally are integrated into the systems by $SI_1$. Note that the blue arrows trace the provenance path of a single chiplet produced by $CM_1$, integrated into an IC by $ICM_1$, and finally used by $SI$. In addition, the background colors indicate whether the corresponding blockchain is trusted (green) or untrusted (red). A key goal of our scheme is to establish a trust measure applicable to ICs whose provenance graph crosses trust boundaries similar to the one shown.

\subsection{Trusted Authority (TA)}
We assume the presence of one or more Trusted Authorities (\textit{TA}) within each blockchain network. A \textit{TA} can be a governmental agency or a group consisting of one or more members of the blockchain consortium, designated to make binding and final decisions on reports.
We do not define a specific method for selecting \textit{TAs}; instead, this decision is left to the discretion of the blockchain consortium members.\textit{TAs} are responsible for evaluating reports submitted by part (i.e., chiplets or ICs) manufacturers, and \textit{SIs} or \textit{EUs}. Upon receiving a report, one of the \textit{TAs} validates the authenticity of the report that contains information of chiplets or ICs that are faulty, counterfeit, or compromised. We can collectively refer to it as ``defective".

\subsection{Chiplet and IC Life Cycles}
Our reputation scheme differentiates between the life cycles of chiplets and ICs. The life cycle of a chiplet begins with its manufacturer and ends with an IC manufacturer, whereas an IC's life cycle starts at its manufacturer and ends with a SI or EU. There are two primary reasons for this separation. Firstly, we conservatively wait until a verification process is completed before we reward or penalize the reputation of the entities involved. These verification processes occur at IC manufacturers for chiplets and at SIs/EUs for ICs. IC manufacturers are expected to check the quality of chiplets before incorporating them into IC fabrication. Similarly, SIs/EUs verify ICs through intensive testing or manual use. Based on these findings, we can reward or penalize the chiplet or IC manufacturers and their distributors. The second reason is that the time interval between a chiplet's fabrication and its incorporation into an IC, which then reaches the SI/EU, can span months or even years. Dividing this interval into two stages helps shorten the feedback delay between transactions and reputation updates.

\subsection{Cross-blockchain transactions} \label{sec:reputation/crossbc}
Transactions that cross trust boundaries carry high risk. Our reputation scheme assigns this risk to the buyer who is responsible for a product to cross from an untrusted blockchain (where the seller belongs) to the trusted blockchain. We achieve this by considering the buyer as a member of the untrusted blockchain for the purpose of calculating reputation penalties associated with such transactions. The exact mechanism of how this is done is described in~\Cref{sec:reputation/scheme}.

Furthermore, to identify chains that frequently sell defective products, it is useful to assign \emph{chain reputation} for each blockchain that a given chain transacts with. We achieve this by introducing \emph{meta-entities} that signify cross-blockchain trust. Let $\mathtt{UB}$ and $\mathtt{TB}$ be an untrusted and trusted blockchain, respectively. Then, the meta-entity $\mathtt{X^{UB}_{TB}}$ acts as a logical intermediary for transactions between a seller ($S$) in $\mathtt{UB}$ and a buyer ($B$) in $\mathtt{TB}$. Thus, a transaction $\mathtt{(S,B,amt)}$ is split into $\mathtt{(S,X^{UB}_{TB},amt)}$ and $\mathtt{(X^{UB}_{TB},B,amt)}$. The reputation calculation functions consider meta-entities like any other entity, and the reputation of $\mathtt{X^{UB}_{TB}}$ can be used for purposes such as making administrative decisions.

\subsection{Reputation Scheme} \label{sec:reputation/scheme}
Our proposed reputation scheme follows two basic principles:

\begin{enumerate}
    \item \emph{Purchase equals endorsement:} We consider a sale by entity $S$ to entity $B$ for an amount of $d$ dollars as an endorsement of $S$ by $B$ for a magnitude of $d$ units.

    \item \emph{Reward conservatively, penalize liberally:} We base the reputation gain/loss calculations based on the additive increase, multiplicative decrease (\textit{AIMD}) approach.
\end{enumerate}

The reputation update procedure begins upon the receipt of a chiplet or IC by an IC manufacturer or SI/EU (hereafter collectively termed as \emph{verifier}) respectively. After inspecting and running checks on the product, the verifier themselves, or through a third party, invokes the $\mathtt{Report()}$ smart contract call. $\mathtt{Report()}$ takes a binary $\mathtt{result}$ flag as input which denotes the outcome of the verification process, with $\mathtt{0}$=pass and $\mathtt{1}$=fail. Naturally, the former results in reputation gain, whereas the latter potentially causes reputation loss. We now separately describe each process in detail. 

\subsubsection{Reputation Gain}
If $\mathtt{result=0}$, the $\mathtt{Report()}$ call internally invokes $\mathtt{RewardReputation()}$ which is responsible for distributing the reputation rewards. To do this, $\mathtt{RewardReputation()}$ first walks through the DAG (see~\Cref{fig:DAG}) and retrieves the trace associated with the product. Each edge is represented as a 3-tuple of the form $\mathtt{(buyer, seller, amount)}$.
Then, for each edge, the seller is rewarded with a reputation gain which is equal to the value of $\mathtt{amount}$ and shown as follows:

\vspace{-10px}
\begin{equation}
    \mathtt{seller.reputation ~+=~ cur\_convert(amount)}
    \label{eq:rep-gain}
\end{equation}

\noindent where, $\mathtt{cur\_convert}$ is a utility function that converts the sale amount to a standard currency for uniformity.

We choose cost as the positive reward metric because it is both simple and objective, while effectively reflecting real-world business and economic dynamics. Higher IC complexity or demand typically leads to increased per-chiplet/IC costs, while deeper supply chains drive costs up due to cumulative markups. Although alternative metrics may be useful in specific contexts, we consider sale amount to be the most broadly applicable and easily quantifiable metric for our framework.

\subsubsection{Reputation Loss}
If $\mathtt{result=1}$, the $\mathtt{Report()}$ call flags the transaction for review by the \textit{TA}. Note that the procedure by which the \textit{TA} validates or rejects a report is beyond the scope of this work. Instead, we consider only the outcome of this process (which is recorded as a transaction on the blockchain), namely, zero or more chiplets/ICs judged to be defective. Although zero chiplets/ICs judged as defective implies that the report was rejected by the TA, it is nevertheless recorded in the blockchain so that IC manufacturers and SI/EUs that frequently raise false alarms can be identified.

The reputation reduction procedure works on a per-defective chiplet/IC basis, i.e., $\mathtt{PenalizeReputation()}$ is invoked for each chiplet/IC found to be defective.
The penalization process for a particular chiplet/IC starts from the manufacturer and traverses the provenance path to the IC manufacturer or SI/EU. For each seller along the path, the fraction of reputation to be reduced is determined based on a \emph{multiplicative decrease parameter} $m$, as well as a \emph{discounting parameter} $d$. The multiplicative decrease parameter $m>1$ is a global parameter defined at the blockchain level that determines the base reputation penalty rate.
The reputation of the chiplet/IC manufacturer is reduced by a factor of $m$, and described as follows:

\vspace{-10px}
\begin{equation}
    \mathtt{manufacturer.reputation ~/=~ m}
    \label{eq:rep-loss-manufacturer}
\end{equation}

The discounting parameter is a per-blockchain parameter used to calculate transitive trust penalties for the remaining sellers along the provenance path. Its value depends on whether the blockchain in which a given transaction took place is trusted ($d_{t}$) or untrusted ($d_{ut}$). $d_{ut}$ is set to 1 so that every member of the sub-path in the untrusted blockchain is penalized equally. This is a necessity since untrusted blockchains may have sybils, i.e., a single entity assuming multiple identities. Furthermore, transactions crossing from an untrusted blockchain to a trusted blockchain also use $d_{ut}$ so that the buyer responsible is penalized at a higher rate to account for the increased risk they introduce by allowing parts to cross blockchain trust boundaries (as discussed in~\Cref{sec:reputation/crossbc}). $d_{t}$ can be set higher than 1 so that entities appearing later in a provenance path are penalized less. Thus, the reputation of the remaining sellers is reduced according to the following expression:

\vspace{-10px}
\begin{equation}
    \mathtt{seller.reputation ~/=~ seller.penalty}
    \label{eq:rep-loss-seller-1}
\end{equation}

\noindent where,

\vspace{-10px}
\begin{equation}
    \mathtt{seller.penalty = parent\_txn.seller.penalty / d}
    \label{eq:rep-loss-seller-2}
\end{equation}

To make the above concepts clear, let us consider~\Cref{fig:DAG} where a chiplet produced by $CM_1$ was judged defective following a report by SI/EU.
By inspecting the blue arrows, we can see that $CM_1$, $CD_1$ and $CD_3$ are penalized equally, whereas $ICM_1$, $ICD_1$, $ICD_2$ are penalized with a discounting factor of $d_{t}$ (denoted as $d$ for brevity). Hence, the reputations of $CM_1$, $CD_1$ and $CD_3$ are slashed by a factor of $m$, and the reputations of $ICM_1$, $ICD_1$ and $ICD_2$ are slashed by $m/d$ and $m/d^2$ and $m/d^3$, respectively.

The above example also illustrates the effect of cross-blockchain transactions:
for the purpose of calculating the penalty, $CD_3$ is essentially considered as a member of the untrusted blockchain (red) even though it actually belongs to the trusted blockchain (green). This is done to emphasize the risk assigned to the entity that is responsible for transacting with the untrusted blockchain. This way, the reputation scheme can capture the high risk of such transactions while keeping the computations simple.

\subsection{Normalized Reputation Score} \label{sec:reputation/norm}
It is important to note that our reputation scheme assigns absolute reputation scores to entities, reflecting their overall trustworthiness based on their past actions and interactions. However, moving to a normalized reputation score is essential, which could provide additional benefits. For example, a normalized score within the range $[0,1]$ could offer a more standardized and easily comparable metric across different entities in the supply chain. 

Let $r$ be the absolute reputation score of an entity $\mathtt{e}$, and $r'$ be its \emph{ideal} absolute reputation score, i.e., what $r$ would have been if $\mathtt{e}$ sold no defective parts.
Clearly, $r'\geq r$.

We define the normalized reputation score of $\mathtt{e}$ as
\begin{equation}
    \begin{cases} 
        1 & r'=0 \\
        \frac{r}{r'} & r'>0 \\
    \end{cases}
    \label{eq:normalized-rep}
\end{equation}

Keeping track of the normalized reputation score of an entity is efficient in terms of both computation and memory complexity, since it only requires $O(1)$ additional space and $O(n)$ additional reward computations (where $n$ is the number of transactions).

\section{Evaluation} \label{sec:eval}
In this section, we demonstrate our proposed reputation scheme with a reference blockchain implementation. We follow this up with a detailed analysis of the increase or decrease of reputation scores across participating members based on their performance in the chain.

\begin{figure}[tb]
    \centering
    \includegraphics[width=1\linewidth]{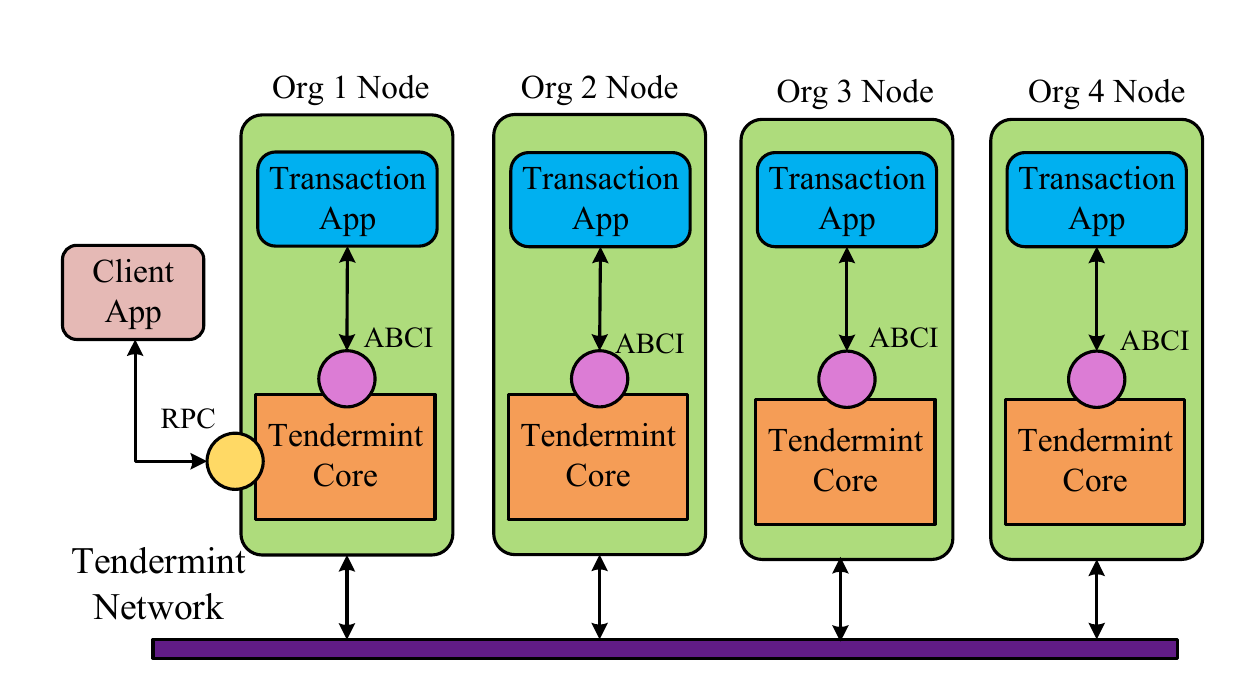}
    \caption{Proposed blockchain-based framework implemented with Tendermint.} \vspace{-10px}
    \label{fig:blockchain_setup}
\end{figure}

\subsection{Blockchain Implementation}
We have implemented our proposed framework using Tendermint~\cite{buchman2016tendermint}. However, it can be extended to other permissioned blockchain frameworks such as Hyperledger `fabric\cite{androulaki2018hyperledger}. \Cref{fig:blockchain_setup} demonstrates an overall representation of our Tendermint architecture. The Tendermint Core performs Byzantine fault tolerant(BFT) state machine replication, which is used to create and manage the blockchain network. Tendermint's ABCI interface accounts for a more modular structure, facilitating the communication between the blockchain and the application's operational logic. This ultimately ensures that the application can process transactions without interfering with the underlying consensus mechanism.

Our Tendermint application consists of a Trusted IC supply chain and a Python script that acts as the client application for sending and receiving transactions to and from the blockchain server. The client application ensures that the data sent over to the blockchain is formatted correctly and the received data is interpreted according to our requirements. The transactions are submitted to the blockchain network via HTTP requests to the Tendermint node's RPC interface\cite{cason2021design}. Each transaction is appended with the user's unique organization ID and their specific role within the supply chain, such as chiplet/IC manufacturer, distributor, or system integrator. As we have demonstrated a proof of concept of our framework, in a real-world implementation, we anticipate that users will be equipped with identity certificates that authenticate their transactions with unique identities and roles within the chain.  These are then used to verify access control before invoking certain operations in our operational logic. Once the transaction undergoes the consensus mechanism and is processed according to our Transaction Application logic, as shown in~\Cref{fig:blockchain}, it is stored immutably in our ledger. The implementation of the $\mathtt{Report()}$ function further populates our ledger with the reputation scores of individual organizations of the blockchain consortiums that can later be queried by organizations to make an informed decision before transacting with the respective organizations. To demonstrate a multi-computer network set-up, we have deployed our Tendermint nodes in 4 docker containers. These validator nodes are configured to connect to their replicated transaction applications, which also run in their own Docker containers. 

\begin{figure*}[tb]
    \centering
    \includegraphics[width=0.9\linewidth]{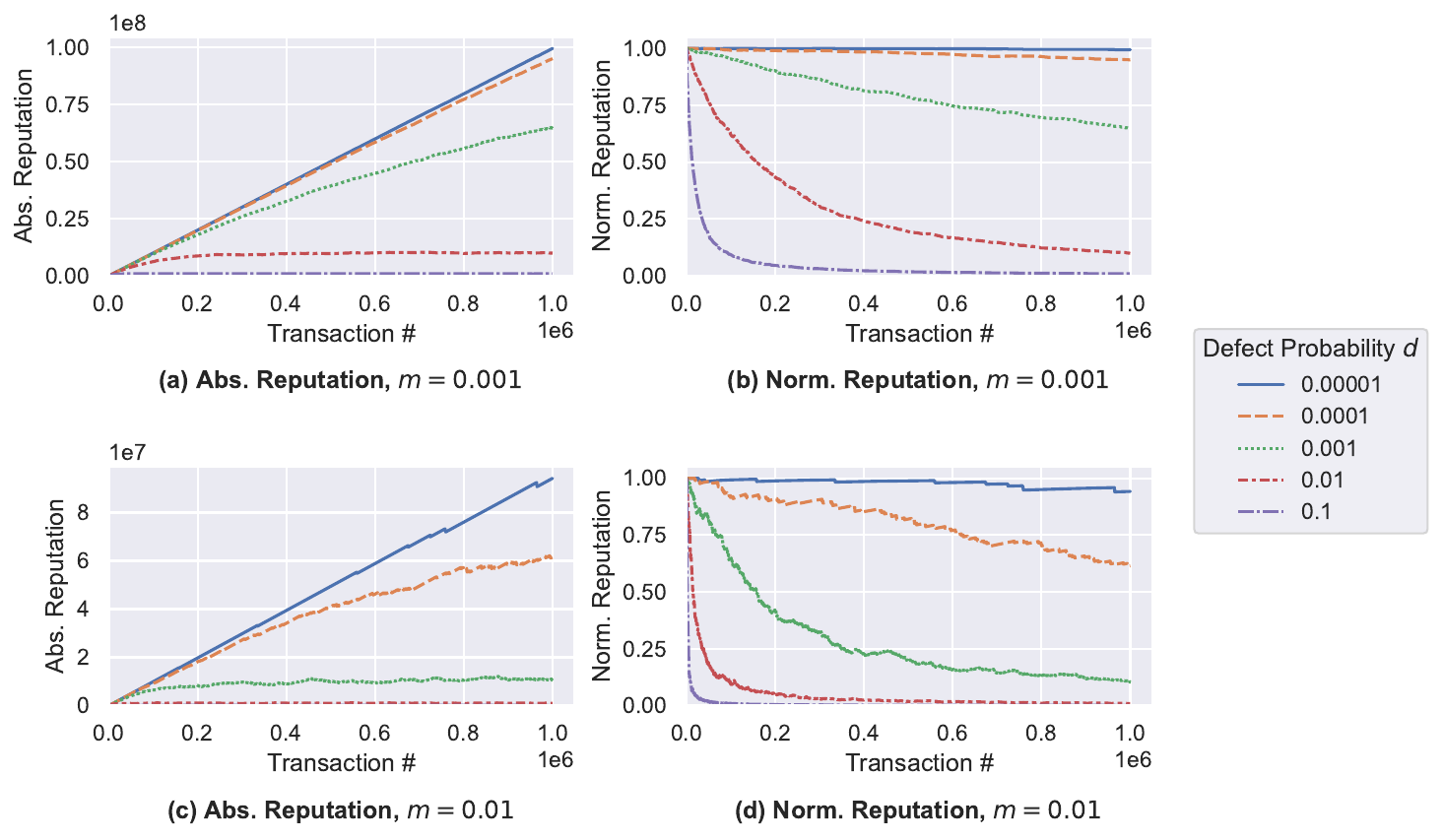}
    \caption{
        Evolution of absolute and normalized reputation with the number of transactions for different defect probabilities ($d$) and multiplicative decrease factors ($m$).
        Figure (a) \& (c) show the absolute reputation that starts at $0$ and increases/decreases according to~\Cref{eq:rep-gain,eq:rep-loss-manufacturer,eq:rep-loss-seller-1,eq:rep-loss-seller-2}. Figure (b) \& (d) show the normalized reputation that starts at $1.0$ and can only decrease, as per~\Cref{eq:normalized-rep}.
    } \vspace{-10px}
    \label{fig:sim-rep-vs-defective-probability}
\end{figure*}

\subsection{Basic Simulation} \label{sec:eval/basic-sim}
The objectives of our initial evaluation were twofold: (1) to analyze the impact of different system parameters on the proposed reputation scheme, and (2) to assess the suitability of the scheme for real-world scenarios.
Among the two system parameters, we focused on the multiplicative decrease parameter $m$, as the choice of discounting factor $d_{ut}$ is more of an administrative decision.
Towards this goal, we simulated how the reputation of a chiplet manufacturer evolves with the number of transactions, considering different values of \emph{defect probability} $d$ (i.e., the likelihood that a chiplet produced by the manufacturer is defective) and the multiplicative decrease factor $m$.
The results of this simulation are shown in~\Cref{fig:sim-rep-vs-defective-probability}.

Several key observations can be drawn from the figure.
First, entities with very low defect probabilities maintain a normalized reputation close to $1$, even after $10^6$ transactions. However, this value is influenced by $m$ -- a higher $m$ leads to a more rapid decline in normalized reputation for the same defect probability.
For instance, for $m=0.001$, an entity with a defect probability of $0.00001$ experiences almost no reputation loss over the course of the simulation.
However, when $m=0.01$, there is a slight but noticeable decline in reputation, similar to what happens with a defect probability of $0.0001$ when $m=0.001$.
\emph{This suggests that $m$ should be selected so that an entity with a defect probability no worse than the typical industry fabrication defect rate experiences almost no loss in normalized reputation.}

Secondly, it is evident that the higher the $d$ value of an entity, the more rapidly its normalized reputation declines.
For example, when $m=0.01$, an entity with $d=0.0001$ experiences a nearly linear decline in normalized reputation to $0.6$ over $10^6$ transactions.
In contrast, an entity with $d=0.01$ or higher is heavily penalized, with its normalized reputation dropping sharply to nearly zero within a few thousand transactions.
Since this observation aligns with the expected behavior of a reputation scheme, it qualitatively supports the suitability of our proposed scheme for real-world applications.
Next, we further investigate this claim through an end-to-end simulation.

\subsection{End-to-end Simulation}
To validate the real-world applicability of the proposed reputation scheme, we developed a simulation platform for the entire semiconductor supply chain using Python.
The platform consists of two components: a supply chain simulator and a reputation simulator.
We now briefly describe each component and explain how they work together to achieve an end-to-end simulation.

\subsubsection{Supply Chain Simulator}
The purpose of the supply chain simulator is to generate a sequence of transactions that closely mimic the flow of parts in a real-world semiconductor supply chain.
\Cref{fig:DAG} illustrates a directed acyclic graph (DAG) representing an example supply chain involving various chiplet and IC manufacturers, distributors, and system integrators, described in~\Cref{subsec:DAG}.
The flow of parts between entities is represented in the simulation as a sequence of transactions in the following format:
\begin{equation}
    T_i=\{PT, S_i, D_i, c, n, \{ID_1, ID_2, \ldots ID_n \} \}
\end{equation}
\noindent where, $T_i$, $PT$, $S_i$, $D_i$, $n$, and $ID$ represent $i^{th}$ transaction, part type, source entity, destination entity, cost of part, number of parts and their IDs, respectively.

The supply chain simulator is designed with flexibility in mind, allowing users to configure various parameters such as the number of transactions, the number of entities of each type, the number of trusted and untrusted consortiums, resale markup percentages, and defect probability.
While this simulator was developed as part of our end-to-end simulation pipeline, we hope that the research community finds it valuable for other studies involving the semiconductor supply chain.

\subsubsection{Reputation Simulator}
The sequence of transactions generated by the supply chain simulator is then fed into the reputation simulator, which tracks the evolution of both absolute and normalized reputations of entities with each transaction.
By decoupling the reputation calculation from the supply chain simulation, we can test various reputation system parameters and compare the results for a specific supply chain instance.

\begin{figure*}[tb]
    \centering

    \subfloat[Chiplet Manufacturers]{
       \includegraphics[width=\linewidth,trim={0 2cm 0 0},clip]{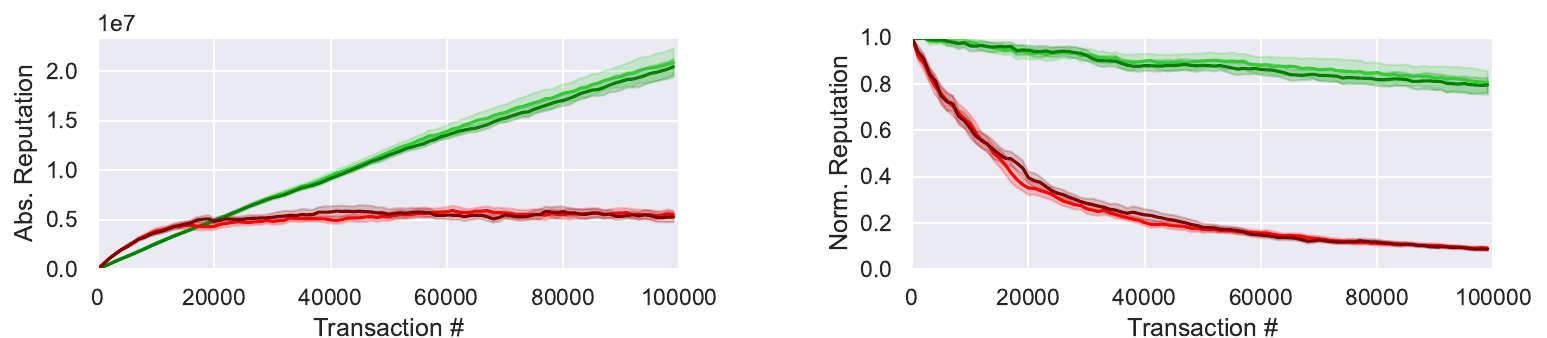}
    }

    \subfloat[Chiplet Distributors]{
       \includegraphics[width=\linewidth]{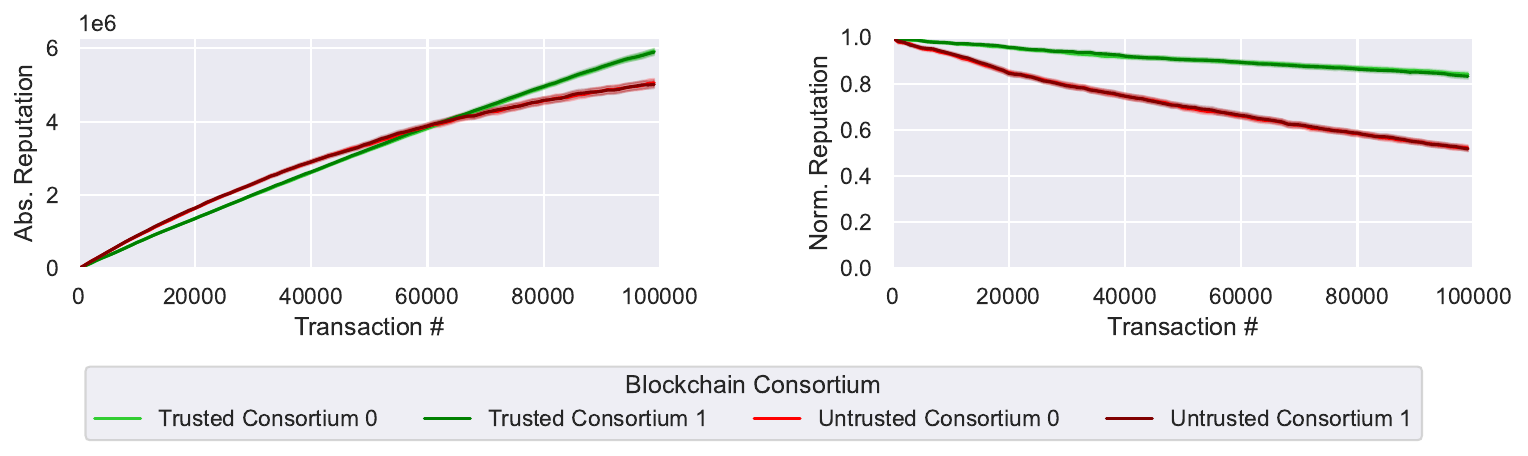}
    }

    \caption{
        Evolution of absolute and normalized reputation of chiplet manufacturers and distributors, aggregated by consortiums, over an end-to-end simulation consisting of $10^6$ transactions. Green and red lines indicate the mean reputation of trusted and untrusted consortiums, respectively. Shaded regions around the lines indicate the $95^{th}$ percentile value. Similar trends are seen for IC manufacturers and distributors.
    } \vspace{-10px}
    \label{fig:sim-end-to-end}
\end{figure*}

We tested the end-to-end simulation pipeline using a sequence of 1 million simulated transactions within a supply chain consisting of 100 chiplet manufacturers, 1000 chiplet distributors, 100 IC manufacturers, and 500 IC distributors.
The results are shown in~\Cref{fig:sim-end-to-end}.
As seen in the figure, the two trusted consortiums consistently outperform the untrusted consortiums in terms of normalized reputation.
This suggests that our proposed reputation scheme is effective for real-world semiconductor supply chains.
However, this analysis focuses on the overall reputation trends of consortiums, not on the behavior of \emph{individual} entities that choose to engage in malicious activities.
We address this question next.

\subsection{Attack Resilience}
\begin{figure*}[tb]
    \centering
    \includegraphics[width=0.9\linewidth]{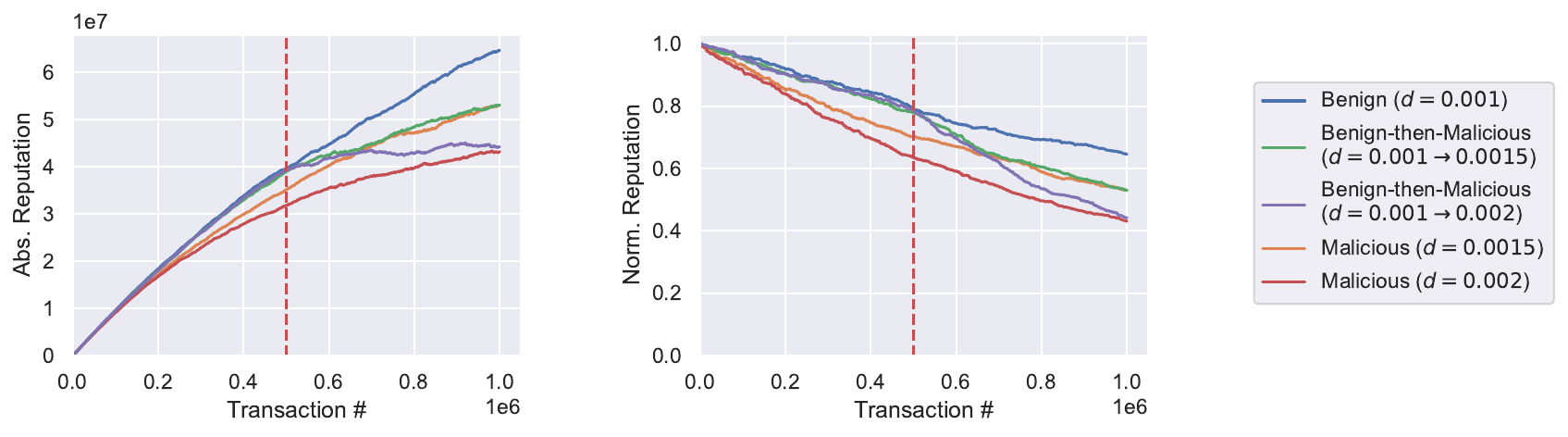}
    \caption{
        Comparison of benign, malicious, and benign-then-malicious behaviors. In this simulation, benign entities are assumed to have a defect probability of $d=0.001$, while malicious entities have a defect probability of either $d=0.0015$ or $d=0.002$. A benign-then-malicious entity behaves benignly for the first 500K transactions (marked by the vertical red dashed line) before switching to malicious behavior.
    }
    \label{fig:sim-rep-free-fall} \vspace{-10px}
\end{figure*}

A major threat to the integrity of reputation systems is ``sleeper agents'' which act benignly until they build up enough reputation, then launch attacks on the system. In poorly designed reputation schemes, these entities may maintain high reputations even after initiating attacks, by virtue of the reputation accrued beforehand. To evaluate the resilience of our proposed reputation scheme against such attacks, we designed a simulation experiment with three types of behaviors: benign, malicious, and benign-then-malicious. Entities exhibiting benign behavior are assumed to have a defect probability of $d=0.001$. We consider two levels of malicious behavior, with defect probabilities of $d=0.0015$ and $d=0.002$. Benign-then-malicious entities behave benignly for the first 500K transactions, then switch to one of the two malicious behavior levels for the next 500K transactions.
The results of this experiment are shown in~\Cref{fig:sim-rep-free-fall}.

In~\Cref{fig:sim-rep-free-fall}, the vertical dashed red line marks the point at which benign-then-malicious entities begin exhibiting malicious behavior. Up until that point, both their absolute and normalized reputations follow a similar trajectory to that of a benign entity. The key observation from the figure is that after a sufficient number of transactions following the onset of malicious behavior, entities exhibiting benign-then-malicious behavior arrive at roughly the same absolute and normalized reputation as if they had been engaging in malicious behavior from the beginning. Thus, we conclude that benign-then-malicious behavior offers no long-term advantage to entities that adopt it. A more detailed analytical investigation of this phenomenon is left for future work.

It is important to note that, while the simulation allows us to know whether a user is malicious (since we predefine them), a real-world deployed system lacks this knowledge. The only actionable metric available to such a system is the defect probability $d$. Our simulations demonstrate that when the defect probability increases, the reputation system eventually recovers. Moreover, the recovery time can be shortened by increasing the multiplicative decrease factor $m$, enabling the system to identify malicious entities more quickly. Together, these results highlight the system's robustness against long-term risks and its ability to mitigate short-term risks through parameter tuning.

We are not aware of any real-world datasets currently available to evaluate the feasibility of our proposed scheme beyond the insights gained from our extensive simulations. Consequently, we leave the analysis using real-world datasets as a direction for future work.

\section{Discussion} \label{sec:discussions}
In the previous sections, we introduced and evaluated a novel reputation scheme for entities within the global semiconductor supply chain. Unlike most reputation schemes, our proposed approach allows for a unified reputation metric across multiple blockchain consortiums, while permitting variations in how the reputation scheme is implemented within each consortium. As a result, each consortium will have its own perspective on the reputation of a given entity. It is even possible that a Trusted Authority recognized in one consortium may not be regarded as trusted in another. While this approach might seem counterintuitive, we believe that such flexibility is essential for entities operating in different countries with diverse (and sometimes conflicting) goals to collaborate within a shared blockchain-based provenance system.

We argue that our scheme has a minimal entry barrier in terms of effort and cost, as each consortium is solely responsible for managing transactions involving its own members, and the reputation calculations are computationally lightweight. Moreover, once the infrastructure is established, new entities can integrate seamlessly and benefit from the system without requiring substantial investment.

It is worthwhile to briefly discuss how our proposed scheme handles some common challenges faced by reputation systems. One such challenge is the \emph{cold-start} problem, where an entity that joins a reputation scheme long after its establishment struggles to catch up with entities with longer histories. Our scheme partially mitigates this issue by proposing a normalized reputation. Another challenge is \emph{wash trading}, where a group of entities artificially inflates their sales and reputations by repeatedly buying and selling from each other. Our scheme does not directly address wash trading, and enhancing it to resist such manipulation is left for future work. A third issue is \emph{buyer tampering}, where a buyer intentionally alters a purchased product to damage the reputation of the seller.
While solutions like ECID and PUFs may help partially mitigate this, this issue falls outside the scope of this work. We leave the evaluation of the resilience of our scheme against these and other sophisticated attacks for future research.

\section{Conclusion} \label{sec:conclusion}
In today's landscape of horizontally integrated semiconductor supply chains and the advent of 2.5D/3D ICs, it has become increasingly difficult for organizations to accurately assess the behavior and reliability of the organizations with which they are conducting their transactions. To mitigate this issue, we have first detailed a blockchain framework for the supply chain provenance of a 2.5/3D IC. Our paper then introduces a novel evaluation parameter that assigns reputation scores to entities based on their performance within the supply chain. This is accomplished through an additive increase and multiplicative decrease model, which dynamically adjusts reputation scores over time; good actors are rewarded with progressively increasing incentives, while bad actors face proportionate penalties, thereby maintaining a balanced and fair system of accountability. Our paper concludes with a threefold comprehensive analysis of the efficacy of our proposed reward scheme. First, we assess the performance of manufacturers under varying percentages of defective chips; second, we evaluate the performance of different participating entities within both trusted and untrusted consortiums; third, we compare the reputation scheme of entities as they behave maliciously at different points in the supply chain. Our proposed model provides a detailed assessment of the reliability of entities within a supply chain. Blockchain members can leverage this data for effective administrative decision-making, ensuring a more transparent and trustworthy supply chain ecosystem.

\section*{Acknowledgments} 
This material is based upon work supported by the Air Force Office of Scientific Research under award number FA9550-23-1-0312. Any opinions, findings, and conclusions, or recommendations expressed in this material are those of the author(s) and do not necessarily reflect the views of the United States Air Force.

\bibliographystyle{ieeetr}
\bibliography{bibliography-uguin,3D-IC}

\end{document}